# Limits on $WWZ$ and $WW\gamma$ couplings from $WW$ and $WZ$ production in $p\overline{p}$ collisions at $\sqrt{s} = 1.8$ TeV

**Abstract**

Direct limits are set on $WWZ$ and $WW\gamma$ three-boson couplings in a search for $WW$ and $WZ$ production with high transverse momentum in $p\overline{p}$ collisions at $\sqrt{s} = 1.8$ TeV, using the Collider Detector at Fermilab. The results are in agreement with the SU(2) $\times$ U(1) model of electroweak interactions. Assuming Standard Model $WW\gamma$ couplings, the limits are interpreted as direct evidence for a non-zero $WWZ$ coupling at subprocess energies near 500 GeV. Alternatively, assuming identical $WWZ$ and $WW\gamma$ couplings, bounds $-0.11 < \kappa < 2.27$ and $-0.81 < \lambda < 0.84$ are obtained at 95% C.L. for a form factor scale 1000 GeV.

*The CDF Collaboration*

F. Abe,[13] M. G. Albrow,[7] S. R. Amendolia,[23] D. Amidei,[16] J. Antos,[28] C. Anway-Wiese,[4] G. Apollinari,[26] H. Areti,[7] M. Atac,[7] P. Auchincloss,[25] F. Azfar,[21] P. Azzi,[20] N. Bacchetta,[20] W. Badgett,[16] M. W. Bailey,[18] J. Bao,[35] P. de Barbaro,[25] A. Barbaro-Galtieri,[14] V. E. Barnes,[24] B. A. Barnett,[12] P. Bartalini,[23] G. Bauer,[15] T. Baumann,[9] F. Bedeschi,[23] S. Behrends,[3] S. Belforte,[23] G. Bellettini,[23] J. Bellinger,[34] D. Benjamin,[31] J. Benlloch,[15] J. Bensinger,[3] D. Benton,[21] A. Beretvas,[7] J. P. Berge,[7] S. Bertolucci,[8] A. Bhatti,[26] K. Biery,[11] M. Binkley,[7] F. Bird,[29] D. Bisello,[20] R. E. Blair,[1] C. Blocker,[3] A. Bodek,[25] W. Bokhari,[15] V. Bolognesi,[23] D. Bortoletto,[24] C. Boswell,[12] T. Boulos,[14] G. Brandenburg,[9]

C. Bromberg,[17] E. Buckley-Geer,[7] H. S. Budd,[25] K. Burkett,[16] G. Busetto,[20]
A. Byon-Wagner,[7] K. L. Byrum,[1] J. Cammerata,[12] C. Campagnari,[7] M. Campbell,[16]
A. Caner,[7] W. Carithers,[14] D. Carlsmith,[34] A. Castro,[20] Y. Cen,[21] F. Cervelli,[23]
H. Y. Chao,[28] J. Chapman,[16] M.-T. Cheng,[28] G. Chiarelli,[23] T. Chikamatsu,[32]
C. N. Chiou,[28] L. Christofek,[10] S. Cihangir,[7] A. G. Clark,[23] M. Cobal,[23]
M. Contreras,[5] J. Conway,[27] J. Cooper,[7] M. Cordelli,[8] C. Couyoumtzelis,[23]
D. Crane,[1] J. D. Cunningham,[3] T. Daniels,[15] F. DeJongh,[7] S. Delchamps,[7]
S. Dell'Agnello,[23] M. Dell'Orso,[23] L. Demortier,[26] B. Denby,[23] M. Deninno,[2]
P. F. Derwent,[16] T. Devlin,[27] M. Dickson,[25] J. R. Dittmann,[6] S. Donati,[23]
R. B. Drucker,[14] A. Dunn,[16] K. Einsweiler,[14] J. E. Elias,[7] R. Ely,[14] E. Engels, Jr.,[22]
S. Eno,[5] D. Errede,[10] S. Errede,[10] Q. Fan,[25] B. Farhat,[15] I. Fiori,[2] B. Flaugher,[7]
G. W. Foster,[7] M. Franklin,[9] M. Frautschi,[18] J. Freeman,[7] J. Friedman,[15] H. Frisch,[5]
A. Fry,[29] T. A. Fuess,[1] Y. Fukui,[13] S. Funaki,[32] G. Gagliardi,[23] S. Galeotti,[23]
M. Gallinaro,[20] A. F. Garfinkel,[24] S. Geer,[7] D. W. Gerdes,[16] P. Giannetti,[23]
N. Giokaris,[26] P. Giromini,[8] L. Gladney,[21] D. Glenzinski,[12] M. Gold,[18] J. Gonzalez,[21]
A. Gordon,[9] A. T. Goshaw,[6] K. Goulianos,[26] H. Grassmann,[6] A. Grewal,[21]
L. Groer,[27] C. Grosso-Pilcher,[5] C. Haber,[14] S. R. Hahn,[7] R. Hamilton,[9]
R. Handler,[34] R. M. Hans,[35] K. Hara,[32] B. Harral,[21] R. M. Harris,[7] S. A. Hauger,[6]
J. Hauser,[4] C. Hawk,[27] J. Heinrich,[21] D. Cronin-Hennessy,[6] R. Hollebeek,[21]
L. Holloway,[10] A. Hölscher,[11] S. Hong,[16] G. Houk,[21] P. Hu,[22] B. T. Huffman,[22]
R. Hughes,[25] P. Hurst,[9] J. Huston,[17] J. Huth,[9] J. Hylen,[7] M. Incagli,[23]
J. Incandela,[7] H. Iso,[32] H. Jensen,[7] C. P. Jessop,[9] U. Joshi,[7] R. W. Kadel,[14]
E. Kajfasz,[7a] T. Kamon,[30] T. Kaneko,[32] D. A. Kardelis,[10] H. Kasha,[35] Y. Kato,[19]
L. Keeble,[8] R. D. Kennedy,[27] R. Kephart,[7] P. Kesten,[14] D. Kestenbaum,[9]
R. M. Keup,[10] H. Keutelian,[7] F. Keyvan,[4] D. H. Kim,[7] H. S. Kim,[11] S. B. Kim,[16]
S. H. Kim,[32] Y. K. Kim,[14] L. Kirsch,[3] P. Koehn,[25] K. Kondo,[32] J. Konigsberg,[9]
S. Kopp,[5] K. Kordas,[11] W. Koska,[7] E. Kovacs,[7a] W. Kowald,[6] M. Krasberg,[16]

J. Kroll,[7] M. Kruse,[24] S. E. Kuhlmann,[1] E. Kuns,[27] A. T. Laasanen,[24] N. Labanca,[23] S. Lammel,[4] J. I. Lamoureux,[3] T. LeCompte,[10] S. Leone,[23] J. D. Lewis,[7] P. Limon,[7] M. Lindgren,[4] T. M. Liss,[10] N. Lockyer,[21] C. Loomis,[27] O. Long,[21] M. Loreti,[20] E. H. Low,[21] J. Lu,[30] D. Lucchesi,[23] C. B. Luchini,[10] P. Lukens,[7] J. Lys,[14] P. Maas,[34] K. Maeshima,[7] A. Maghakian,[26] P. Maksimovic,[15] M. Mangano,[23] J. Mansour,[17] M. Mariotti,[20] J. P. Marriner,[7] A. Martin,[10] J. A. J. Matthews,[18] R. Mattingly,[15] P. McIntyre,[30] P. Melese,[26] A. Menzione,[23] E. Meschi,[23] G. Michail,[9] S. Mikamo,[13] M. Miller,[5] R. Miller,[17] T. Mimashi,[32] S. Miscetti,[8] M. Mishina,[13] H. Mitsushio,[32] S. Miyashita,[32] Y. Morita,[23] S. Moulding,[26] J. Mueller,[27] A. Mukherjee,[7] T. Muller,[4] P. Musgrave,[11] L. F. Nakae,[29] I. Nakano,[32] C. Nelson,[7] D. Neuberger,[4] C. Newman-Holmes,[7] L. Nodulman,[1] S. Ogawa,[32] S. H. Oh,[6] K. E. Ohl,[35] R. Oishi,[32] T. Okusawa,[19] C. Pagliarone,[23] R. Paoletti,[23] V. Papadimitriou,[31] S. Park,[7] J. Patrick,[7] G. Pauletta,[23] M. Paulini,[14] L. Pescara,[20] M. D. Peters,[14] T. J. Phillips,[6] G. Piacentino,[2] M. Pillai,[25] R. Plunkett,[7] L. Pondrom,[34] N. Produit,[14] J. Proudfoot,[1] F. Ptohos,[9] G. Punzi,[23] K. Ragan,[11] F. Rimondi,[2] L. Ristori,[23] M. Roach-Bellino,[33] W. J. Robertson,[6] T. Rodrigo,[7] J. Romano,[5] L. Rosenson,[15] W. K. Sakumoto,[25] D. Saltzberg,[5] A. Sansoni,[8] V. Scarpine,[30] A. Schindler,[14] P. Schlabach,[9] E. E. Schmidt,[7] M. P. Schmidt,[35] O. Schneider,[14] G. F. Sciacca,[23] A. Scribano,[23] S. Segler,[7] S. Seidel,[18] Y. Seiya,[32] G. Sganos,[11] A. Sgolacchia,[2] M. Shapiro,[14] N. M. Shaw,[24] Q. Shen,[24] P. F. Shepard,[22] M. Shimojima,[32] M. Shochet,[5] J. Siegrist,[29] A. Sill,[31] P. Sinervo,[11] P. Singh,[22] J. Skarha,[12] K. Sliwa,[33] D. A. Smith,[23] F. D. Snider,[12] L. Song,[7] T. Song,[16] J. Spalding,[7] L. Spiegel,[7] P. Sphicas,[15] A. Spies,[12] L. Stanco,[20] J. Steele,[34] A. Stefanini,[23] K. Strahl,[11] J. Strait,[7] D. Stuart,[7] G. Sullivan,[5] K. Sumorok,[15] R. L. Swartz, Jr.,[10] T. Takahashi,[19] K. Takikawa,[32] F. Tartarelli,[23] W. Taylor,[11] P. K. Teng,[28] Y. Teramoto,[19] S. Tether,[15] D. Theriot,[7] J. Thomas,[29] T. L. Thomas,[18] R. Thun,[16] M. Timko,[33] P. Tipton,[25] A. Titov,[26] S. Tkaczyk,[7] K. Tollefson,[25] A. Tollestrup,[7]

J. Tonnison,[24] J. F. de Troconiz,[9] J. Tseng,[12] M. Turcotte,[29] N. Turini,[23]
N. Uemura,[32] F. Ukegawa,[21] G. Unal,[21] S. C. van den Brink,[22] S. Vejcik, III,[16]
R. Vidal,[7] M. Vondracek,[10] D. Vucinic,[15] R. G. Wagner,[1] R. L. Wagner,[7]
N. Wainer,[7] R. C. Walker,[25] C. Wang,[6] C. H. Wang,[28] G. Wang,[23] J. Wang,[5]
M. J. Wang,[28] Q. F. Wang,[26] A. Warburton,[11] G. Watts,[25] T. Watts,[27] R. Webb,[30]
C. Wei,[6] C. Wendt,[34] H. Wenzel,[14] W. C. Wester, III,[7] T. Westhusing,[10]
A. B. Wicklund,[1] E. Wicklund,[7] R. Wilkinson,[21] H. H. Williams,[21] P. Wilson,[5]
B. L. Winer,[25] J. Wolinski,[30] D. Y. Wu,[16] X. Wu,[23] J. Wyss,[20] A. Yagil,[7] W. Yao,[14]
K. Yasuoka,[32] Y. Ye,[11] G. P. Yeh,[7] P. Yeh,[28] M. Yin,[6] J. Yoh,[7] C. Yosef,[17]
T. Yoshida,[19] D. Yovanovitch,[7] I. Yu,[35] J. C. Yun,[7] A. Zanetti,[23] F. Zetti,[23]
L. Zhang,[34] S. Zhang,[16] W. Zhang,[21] and S. Zucchelli[2]

(CDF Collaboration)

[1] *Argonne National Laboratory, Argonne, Illinois 60439*

[2] *Istituto Nazionale di Fisica Nucleare, University of Bologna, I-40126 Bologna, Italy*

[3] *Brandeis University, Waltham, Massachusetts 02254*

[4] *University of California at Los Angeles, Los Angeles, California 90024*

[5] *University of Chicago, Chicago, Illinois 60637*

[6] *Duke University, Durham, North Carolina 27708*

[7] *Fermi National Accelerator Laboratory, Batavia, Illinois 60510*

[8] *Laboratori Nazionali di Frascati, Istituto Nazionale di Fisica Nucleare, I-00044 Frascati, Italy*

[9] *Harvard University, Cambridge, Massachusetts 02138*

[10] *University of Illinois, Urbana, Illinois 61801*

[11] *Institute of Particle Physics, McGill University, Montreal H3A 2T8, and University of Toronto, Toronto M5S 1A7, Canada*

[12] *The Johns Hopkins University, Baltimore, Maryland 21218*

[13] *National Laboratory for High Energy Physics (KEK), Tsukuba, Ibaraki 305, Japan*

[14] *Lawrence Berkeley Laboratory, Berkeley, California 94720*

[15] *Massachusetts Institute of Technology, Cambridge, Massachusetts 02139*

[16] *University of Michigan, Ann Arbor, Michigan 48109*

[17] *Michigan State University, East Lansing, Michigan 48824*

[18] *University of New Mexico, Albuquerque, New Mexico 87131*

[19] *Osaka City University, Osaka 588, Japan*

[20] *Universita di Padova, Istituto Nazionale di Fisica Nucleare, Sezione di Padova, I-35131 Padova, Italy*

[21] *University of Pennsylvania, Philadelphia, Pennsylvania 19104*

[22] *University of Pittsburgh, Pittsburgh, Pennsylvania 15260*

[23] *Istituto Nazionale di Fisica Nucleare, University and Scuola Normale Superiore of Pisa, I-56100 Pisa, Italy*

[24] *Purdue University, West Lafayette, Indiana 47907*

[25] *University of Rochester, Rochester, New York 14627*

[26] *Rockefeller University, New York, New York 10021*

[27] *Rutgers University, Piscataway, New Jersey 08854*

[28] *Academia Sinica, Taiwan 11529, Republic of China*

[29] *Superconducting Super Collider Laboratory, Dallas, Texas 75237*

[30] *Texas A&M University, College Station, Texas 77843*

[31] *Texas Tech University, Lubbock, Texas 79409*

[32] *University of Tsukuba, Tsukuba, Ibaraki 305, Japan*

[33] *Tufts University, Medford, Massachusetts 02155*

[34] *University of Wisconsin, Madison, Wisconsin 53706*

[35] *Yale University, New Haven, Connecticut 06511*

Since formulation of the SU(2)×U(1) gauge theory of electroweak interactions[1], many of its predictions have been confirmed, including the existence of the $W$ and $Z$ force carriers. The parameters of the model have been determined with ever increasing precision, but only now are there direct tests of the predicted interactions of $W$, $Z$, and $\gamma$ bosons with each other. These interactions are the most characteristic and fundamental signatures of non-Abelian symmetry in the theory. The predicted interactions are described by trilinear couplings $WWZ$ and $WW\gamma$, which we address here, as well as quadrilinear couplings. Boson pair production is sensitive to these couplings[2], and the $WW\gamma$ coupling has been tested directly in the process $p\overline{p} \rightarrow W\gamma$[3]. In this letter we report direct information about the $WWZ$ and $WW\gamma$ couplings obtained from a search for $WW$ and $WZ$ production with large boson transverse momentum ($P_T$) in $p\overline{p}$ collisions at $\sqrt{s} = 1.8$ TeV.

Indirect limits on the $WWZ$ and $WW\gamma$ couplings have previously been set based on one-loop effects at low energies and precision measurements at the $Z$ resonance[4, 5]. The direct measurements provided by diboson production are valuable because they are unambiguous as to their interpretation. They do not require additional theoretical assumptions or calculation of loop diagrams, which can present theoretical ambiguities for non-standard models. Furthermore, they are sensitive to directions in the space of couplings which are not well constrained by the indirect limits.

The most general $WWZ$ and $WW\gamma$ couplings consistent with Lorentz invariance have been formulated and may be parametrized in terms of fourteen independent couplings (or form factors), seven for the $WWZ$ vertex and seven for the $WW\gamma$ vertex[6]. They are usually denoted $g_1^V$, $g_4^V$, $g_5^V$, $\lambda^V$, $\kappa^V$, $\tilde{\lambda}^V$, and $\tilde{\kappa}^V$ where $V$ is either $Z$ (for $WWZ$) or $\gamma$ (for $WW\gamma$). The standard $\mathrm{SU}(2) \times \mathrm{U}(1)$ electroweak theory corresponds to the choice $g_1^\gamma = g_1^Z = 1$ and $\kappa^\gamma = \kappa^Z = 1$ with all other couplings set to zero.

In the Standard Model, the dominant contribution to diboson ($WW$ or $WZ$) production in $p\overline{p}$ collisions at $\sqrt{s} = 1.8\,\mathrm{TeV}$ comes from two types of Feynman diagrams, the t- or u-channel diagrams, which involve the couplings of $W$ and $Z$ to fermions, and the s-channel diagrams which are the only ones containing the three-boson coupling. To the extent that the fermionic couplings of the $W$ and $Z$ have been well tested, we may regard diboson production as primarily a test of the three-boson couplings. There are substantial cancellations between the s-channel and the t- or u-channel diagrams, resulting in cross sections of 9.5 pb and 2.5 pb for $WW$ and $WZ$ production respectively. If any of the three-boson couplings differ from the standard model values then the cancellations are reduced and the cross section increases. The enhancement is greatest at high boson $P_T$ where the strongest cancellations occur in the standard model. Therefore, this analysis looks for anomalously large cross sections at high boson $P_T$ in order to obtain information on the couplings.

The data for the analysis were recorded with the Collider Detector at Fermilab during the 1992-93 Fermilab Tevatron collider run, corresponding to an integrated luminosity of $19.6\,\mathrm{pb}^{-1}$. The detector has been described in detail elsewhere[7]. Here we give a brief description of the components relevant to this analysis. The location of the event vertex is measured along the beam direction with a time projection chamber (VTX). The momenta of charged particles are measured in the central tracking chamber (CTC), which is surrounded by a 1.4 T superconducting solenoidal magnet. Outside the CTC, the calorimeter is organized in electromagnetic (EM) and hadronic (HAD) compartments with projective towers covering the pseudorapidity range $|\eta| \leq 3.6$. Outside the central calorimeter, the region $|\eta| \leq 1.0$ is instrumented with drift chambers for muon identification.

Each electron is identified by an isolated cluster in either the central EM calorimeter ($|\eta| \leq 1.1$) which matches a track in the CTC or the endplug EM calorimeter ($1.1 \leq |\eta| \leq 2.4$) with associated hits in the VTX. Each muon is identified by an isolated track in the CTC with minimum ionizing energy in the calorimeter. Events with one or more muons must have at least one muon with matching hits in the muon chambers. The presence of neutrinos is inferred from missing transverse energy ($\not\!\!E_T$), which is measured by the magnitude of the vector sum of the calorimeter tower energies perpendicular to the beam axis. Jet energy is measured by clustering the EM and HAD calorimeter energy within a cone $\Delta R < 0.4$, where $\Delta R = \sqrt{\Delta\phi^2 + \Delta\eta^2}$, and $\phi$ is the azimuthal angle[8].

We search for $WW$ and $WZ$ event candidates consistent with the decay of one boson to leptons and the other to hadrons. This choice of decay channels gives better sensitivity to anomalous three-boson couplings than the purely leptonic channels because the leptonic branching fractions of the $W$ and $Z$ are small and because the acceptance of the detector for jets is larger than for leptons. Background from the QCD processes $p\overline{p} \to W + \text{jets}$ and $p\overline{p} \to Z + \text{jets}$ is greatly reduced by requiring a large boson $P_T$, while retaining good sensitivity to anomalous three-boson couplings[6]. Background QCD processes are calculated at Born level[9], including simulation of the CDF detector and jet fragmentation using an adaptation of the HERWIG program[10, 11]. The boson $P_T$ requirement for $WW$ and $WZ$ event selection is chosen so that less than one background event is expected in the final sample. With this choice it is unnecessary to perform a background subtraction and any theoretical uncertainty in the background calculation is avoided.

A leptonic $W$ decay is identified by an isolated electron or muon with $P_T > 20\,\mathrm{GeV/c}$ and $\not\!\!E_T > 20\,\mathrm{GeV}$ forming a transverse mass $M_T > 40\,\mathrm{GeV/c^2}$. A leptonic $Z$ decay is identified by an electron or muon pair of opposite charge forming an invariant mass $70 < M < 110\,\mathrm{GeV/c^2}$. In events with a leptonic $W$ or $Z$ decay, a candidate hadronic $W$ or $Z$ decay is defined by the two jets (leading jets) in the event with the highest jet transverse energies ($E_T$). Each jet must have $E_T > 30\,\mathrm{GeV}$ and the invariant mass of the jet pair must be in the range $60 < M_{JJ} < 110\,\mathrm{GeV/c^2}$. The $P_T$ of the two-jet system, interpreted as a hadronic $W$ or $Z$ decay, is required to satisfy $P_T > 130\,\mathrm{GeV/c}$ for leptonic $W$ events or $P_T > 100\,\mathrm{GeV/c}$ for leptonic $Z$ events.

The two-jet mass spectrum is shown in Fig. 1a for events with a leptonic $W$ decay and with both leading jets satisfying $E_T > 30\,\mathrm{GeV}$. The sum of the predicted Standard Model $WW$ and $WZ$ signals plus QCD background is also shown, where the background is normalized to the observed number of W events with two jets minus the predicted signal. Fig. 1b shows the two-jet $P_T$ distribution in the subset of events which satisfy the two-jet mass criterion. The two-jet $P_T$ requirement is indicated by the arrow. One event passes this cut. For events with a leptonic Z decay there are no events which satisfy all selection criteria.

The limits on the couplings follow from a Monte Carlo calculation of expected event yields for various values of the couplings. The Monte Carlo event generator[6, 12] calculates to leading order the processes $p\overline{p} \to W^+W^-$ and $p\overline{p} \to WZ$ with subsequent decay of a $W$ to $e\nu$, $\mu\nu$, or $jj$ and a $Z$ to $ee$, $\mu\mu$, or $jj$. Higher order QCD corrections to the cross section are accounted for by a "K-factor" of $K = 1 + \frac{8}{9}\pi\alpha_s$[6]. MTB2 structure functions are used[13]. Initial and final state QCD radiation effects and jet fragmentation are modelled with an adaptation of HERWIG[10, 11]. The event generator is combined with a detector simulation which includes trigger efficiencies, lepton identification efficiencies, and jet response modeling. A fast parametrization of the full detector simulation was also employed. The trigger and lepton identification efficiencies are determined from the data and amount to 78% for electrons and 79% for muons. The modeling of the jet response and resolution are tuned to agree with studies of collider and test beam data[14]. The two-jet mass resolution is expected to be 9 GeV/$c^2$ for diboson events that would pass our candidate selection criteria. The efficiency of the two-jet mass cut is 88% for events passing all other cuts.

The systematic uncertainties on the yield are the uncertainties in the structure functions (6%), jet $E_T$ scale and resolution (16%), luminosity (4%), lepton identification efficiency (1%), and trigger efficiency (1%). The Monte Carlo acceptance modeling has 3% statistical uncertainty, and a 5% systematic uncertainty allows for differences between fast and full detector simulations. In addition a 14% uncertainty is assigned for the effects of higher order QCD corrections[10, 15, 16]. These uncertainties are combined in quadrature.

The acceptance is a strong function of the couplings, because of the boson $P_T$ cut in combination with a varying boson $P_T$ distribution. For standard model couplings, 0.13 $WW/WZ \to l\nu jj$ events and 0.02 $WZ \to lljj$ events are expected to pass the selection criteria, where $l$ is either an electron or a muon. The observation of one event in the $l\nu jj$ channel and zero events in the $lljj$ channel is therefore not indicative of a departure from standard model couplings, even without consideration of the QCD background.

The predicted yield of high $P_T$ boson pairs is a quadratic function of the anomalous couplings. The lack of an excess of events therefore results in bounds on the couplings which take the form of ellipses in the plane of any two couplings. Since the one event passing all selection criteria could be either signal or background, we calculate the confidence limits from the probability of observing one or less signal events. We do not perform a background subtraction and therefore obtain conservative limits. The probability distribution used is the convolution of a Poisson distribution with a Gaussian, where the Gaussian smears the mean of the Poisson distribution around the expected yield within the systematic uncertainty.

In the calculation of $WW$ and $WZ$ cross sections, the anomalous parts of the couplings are suppressed at high subprocess center of mass energy ($\sqrt{\hat{s}}$) by a dipole form factor[6]:

$$\xi(\hat{s}) = \xi_{SM} + \frac{\xi(0) - \xi_{SM}}{(1 + \hat{s}/\Lambda_{FF}^2)^2} \quad (1)$$

where $\xi$ stands for any of the couplings, $\xi_{SM}$ is its value in the standard model, and $\Lambda_{FF}$ represents the energy scale of unknown phenomena. Without this suppression, the anomalous couplings would result in cross sections that violate unitarity at large $\hat{s}$. With the suppression, the couplings approach their standard model values at energies above the scale $\Lambda_{FF}$, and the cross sections respect unitarity as long as the anomalous couplings are not too large[17].

In Fig. 2 we present bounds on four pairs of couplings. Except as noted in the figure caption, for each case all the other couplings are fixed at the standard model values. Each pair is constrained to the interior of an ellipse, which is a two dimensional section through an ellipsoidal allowed region in the fourteen dimensional space of three boson couplings. Because the bosons are required to have high $P_T$ our search is most sensitive to the couplings at energies near $\sqrt{\hat{s}} = 500\,\mathrm{GeV}$. The limit contours, however, correspond to the value of the couplings at $\sqrt{\hat{s}} = 0$ and therefore depend on the choice of $\Lambda_{FF}$ according to equation (1). The bounds are shown for $\Lambda_{FF} = 1000$ GeV and $\Lambda_{FF} = 1500$ GeV. The unitarity bounds, which depend strongly on $\Lambda_{FF}$, are also shown[17, 18]. For values of $\Lambda_{FF}$ larger than about 1600 GeV the bounds from unitarity are stronger than the bounds from the search.

Fig. 2a shows limits in the plane $\lambda^\gamma$ *vs.* $\lambda^Z$. The limits are stronger for $\lambda^Z$, illustrating the fact that the search is in general more sensitive to the $WWZ$ couplings. It is therefore complementary to studies of the process $p\bar{p} \to W\gamma$[3].

The limits of Fig. 2b focus on the $WWZ$ vertex, assuming that the $WW\gamma$ couplings take their standard model values. Bounds are shown for the couplings $g_1^Z$ and $\kappa^Z$, which are the only $WWZ$ couplings predicted to be nonzero in the standard model. The fact that the point $g_1^Z = \kappa^Z = 0$ lies outside the allowed region can be interpreted as direct evidence for a non-zero WWZ coupling, and for the resulting destructive interference between s-channel and t- or u-channel diagrams which takes place in the standard model. Specifically, the search is directly sensitive to the WWZ coupling in the region $\sqrt{\hat{s}} = 500\,\mathrm{GeV}$. If the WWZ coupling were zero in this region, the s-channel diagram containing the WWZ vertex would not contribute to the am-plitude, and the other diagrams by themselves would predict the observation of $15 \pm 3$ events, where the uncertainty is systematic. Independent of the choice of $\Lambda_{FF}$, this possibility is excluded at greater than 99% CL.

Figs. 2c and 2d show limits on the couplings $\kappa$ and $\lambda$, assuming specific relations between the $WWZ$ and $WW\gamma$ couplings. In Fig. 2c, the $WWZ$ couplings are assumed to equal the $WW\gamma$ couplings. The resulting 95% CL limits on $\kappa$ and $\lambda$ separately, assuming that only one departs from its standard model value, are $-0.11 < \kappa < 2.27$ and $-0.81 < \lambda < 0.84$ for the choice $\Lambda_{FF} = 1000\,\mathrm{GeV}$. With the assumption of matching $WWZ$ and $WW\gamma$ couplings, limits also result for the W boson electric quadrupole moment $Q_e^W = \frac{-e}{{M_W}^2}(\kappa - \lambda)$ and magnetic dipole moment $\mu^W = \frac{e}{2M_W}(1 + \kappa + \lambda)$. In the standard model, these moments take the values $Q_e^W = \frac{-e}{{M_W}^2}$ and $\mu^W = \frac{e}{M_W}$. The point $Q_e^W = \mu^W = 0$ is outside the allowed region. Assuming only one of the moments departs from its standard model value, the limits at 95% CL are $-2.42 < Q_e^W/(e/{M_W}^2) < 0.35$ and $0.37 < \mu^W/(e/M_W) < 1.70$ for $\Lambda_{FF} = 1000\,\mathrm{GeV}$.

For Fig. 2d, the relation assumed between the $WWZ$ and $WW\gamma$ couplings is given by the HISZ equations[4], which specify $\lambda^Z$, $\kappa^Z$, and $g_1^Z$ in terms of the independent variables $\kappa^\gamma$ and $\lambda^\gamma$. This prescription preserves SU(2) $\times$ U(1) gauge invariance and is well motivated in an effective Lagrangian approach. The corresponding subspace of anomalous couplings is not well constrained by previous indirect measurements[4]. The individual 95% CL bounds on $\lambda^\gamma$ and $\kappa^\gamma$ are $-0.35 < \kappa^\gamma < 2.57$ and $-0.85 < \lambda^\gamma < 0.81$ for $\Lambda_{FF} = 1000\,\mathrm{GeV}$, if only one of the two is varied from its standard model value.

We thank U. Baur, T. Han and D. Zeppenfeld for Monte Carlo programs and for many stimulating discussions. We thank the Fermilab staff and the technical staffs of the participating institutions for their vital contributions. This work was supported by the U.S. Department of Energy and National Science Foundation; the Italian Istituto Nazionale di Fisica Nucleare; the Ministry of Education, Science and Culture of Japan; the Natural Sciences and Engineering Research Council of Canada; the National Science Council of the Republic of China; the A. P. Sloan Foundation; and the Alexander von Humboldt-Stiftung.

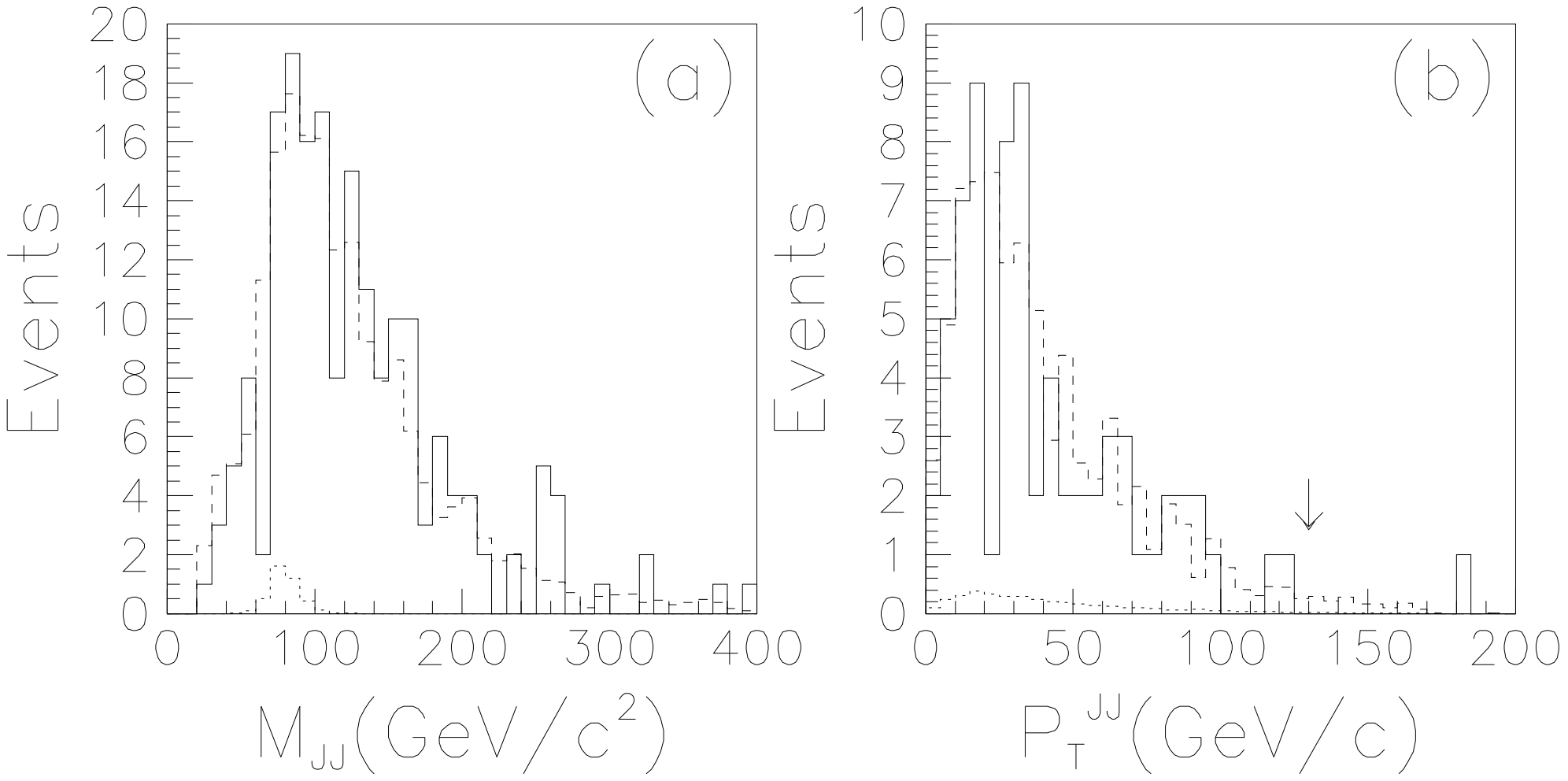

Figure 1: Selection of $WW/WZ \to l\nu jj$ candidates. All event selection cuts except the two-jet mass and two-jet $P_T$ cuts were used to select the events in (a). The subset of events from (a) passing the two-jet mass cut is shown in (b). One event remains after making all cuts. The solid line shows the data, the dots show the predicted standard model diboson signal, and the dashes show the predicted signal plus background shape.

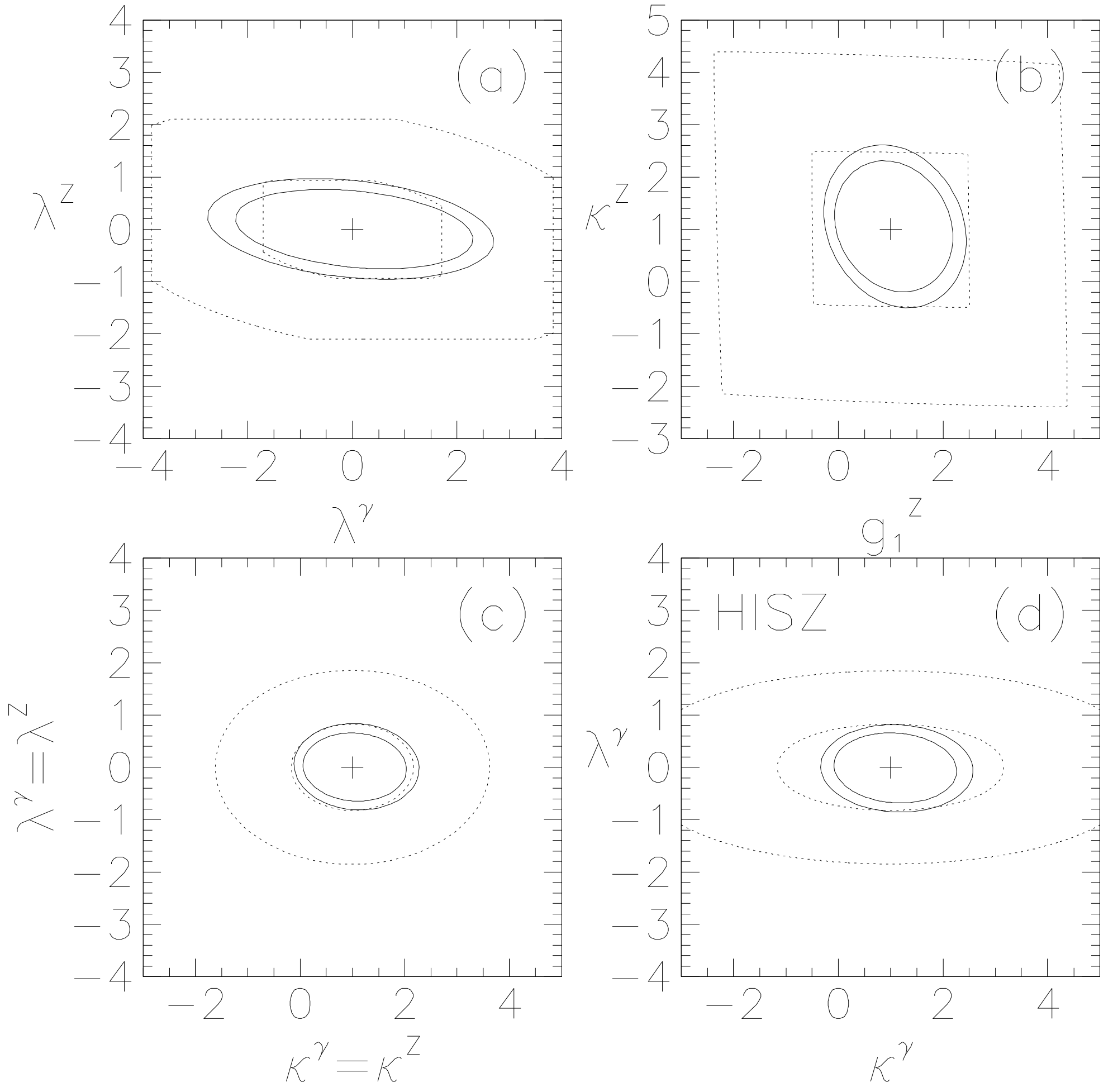

Figure 2: Allowed regions for pairs of anomalous couplings. All couplings, other than those listed for each contour, are held at their standard model values. The solid lines are the 95% CL limits and the dotted lines are the unitarity limits; each is shown for $\Lambda_{FF}$ = 1000 GeV (outer) and 1500 GeV (inner). The + signs indicate the Standard Model values of the couplings. (a) $\lambda^\gamma$ and $\lambda^Z$; (b) ${g_1}^Z$ and $\kappa^Z$; (c) $\kappa$ and $\lambda$ assuming the $WWZ$ and $WW\gamma$ couplings are the same; (d) $\kappa^\gamma$, $\kappa^Z$, $\lambda^\gamma$, $\lambda^Z$ and $g_1^Z$ in the HISZ prescription (see text), with independent variables $\kappa^\gamma$ and $\lambda^\gamma$.